\documentclass[a4paper]{llncs}
                              \institute{ }
\titlerunning{Compiling TM to SMM}
\usepackage{breakcites}
\usepackage{paralist}
\usepackage{graphicx}
\usepackage{listings}

\author{J.-M. Chauvet}
\date{\today}
\title{Compiling Turing Machines into  \\ Storage Modification Machines}
\begin{document}

\maketitle
\begin{abstract}
It is well known that Schönhage's Storage Modification Machines (SMM) can simulate Turing Machines (TM) since Schönhage's original proof of the Turing completeness of the eponymous machines. We propose a simple transformation of TM into SMM, setting the base for a straightforward TM-to-SMM compiler.
\end{abstract}

\section{Introduction}
\label{sec:orge61e553}
It is well known that Schönhage's Storage Modification Machines \cite{Schoenhage1980} can simulate Turing Machines (TM) \cite{Turing:1936:CNA}. Simulation of small universal TM, or other simple universal models such as Post’s tag systems \cite{CockeMinsky1964} and the cellular automaton Rule 110 \cite{Wolfram2002}, is by now a standard way to prove that a large number of alternate models of computation, including a variety of physically-inspired systems, are computationally universal. In the following, we consider a slightly revised version of Schönhage's Storage Modification Machine (SMM) and offer a simple construction to simulate the behavior of a TM. This construction sets the base for a generic TM-to-SMM compiler.

\section{Images of Computation: Turing Machines and Storage Modification Machines}
\label{sec:org2c41e2c}
\emph{Turing Machines}, masquerading as read-write-head-and-tape mechanic contraptions, are in fact quite mathematical, theoretical models of computation. They are best viewed as abstract machines (metaphorically) manipulating symbols on a strip of \emph{tape} according to a table of rules. In a simplified design, the infinite tape is divided into contiguous cells, each cell having a right- and a left-neighbor, which may or not exhibit a \emph{symbol} (from a given, usually finite, alphabet). The \emph{head} when positioned over a cell can read or write this cell's symbol. Finally the machine is characterized by a \emph{state}, also from a given, usually finite, state-set.

An elementary execution step of such a TM is summarized as follows:

\begin{enumerate}
\item \emph{Read} the current cell's symbol \(\sigma\) (or \texttt{blank}, if no symbol is present).
\end{enumerate}
According then to the combination of current state and symbol in the transition table, \((S,\sigma)\):
\begin{enumerate}
\setcounter{enumi}{1}
\item \emph{Write} back a symbol \(\sigma^{\prime}\) on the current cell.
\item \emph{Move} head to one of the two neighbors of the current cell.
\item \emph{Change} internal state to \(S^{\prime}\).
\end{enumerate}

The behavior of this simple TM is entirely specified by its transition table which states, for each relevant \((S,\sigma)\) combination, the data required to perform steps 2, 3 and 4 above. 

In our construction, this data is simply a string which concatenates the symbol  \(\sigma^{\prime}\), the constant \(L\) (left) or \(R\) (right) to indicate which contiguous neighbor to target in step 3, and finally the new state \(S^{\prime}\) of the TM.

Running a TM involves (i) positioning the head on a tape with some initial symbols in some cells, (ii) setting some state as the initial one, and (iii) start consecutively executing the above steps. By convention when no combination of state and symbol is found in the transition table, the machine halts\footnote{The whole point of Turing in his original 1936 paper was to provide a mathematical description of a very simple device capable of arbitrary computations, and prove properties of computation in general -- and in particular, the uncomputability of the \emph{Entscheidungsproblem} (decision problem) of predicting whether a so specified TM would halt or move forever, reading and writing symbols.}.

\emph{Storage Modification Machines} presented here are from a variant in  \cite{Guerraoui2009} where they are used to implement population protocol models. Like a TM, a SMM represents a single computing agent. It is equipped with memory and a processing unit. Its memory stores a finite directed graph of equal out-degree nodes \cite{Schoenhage1980}, with a distinguished node called the \emph{center}. (Edges of this graph are also called \emph{pointers}.) Edges leading out of each node are uniquely labelled by distinct \emph{directions}, drawn from a finite set \(D\).

Any string \(x \in D^*\) refers to the node \(p(x)\) reached from the center by following the sequence of directions labelled by \(x\). (Note that this is somehow similar to moving the head in a TM.) In the variant used here, nodes may have different out-degrees, and we set \(p(x) = \emptyset\) when \(x\) is not a valid path in the graph. 

SMM are furthermore characterized by a \emph{program}, or \emph{control list}, which is a finite list of consecutively numbered instructions (reminiscent of the transition tables in TM). The restricted instruction set is as follows:

\begin{itemize}
\item \emph{\textbf{new} label} creates a new labelled node and makes it the center, setting all its outgoing edges to the previous center.
\item \emph{\textbf{set} xd \textbf{to} y} where \(x,y\) are paths in \(D^*\) and \(d \in D\) is a direction, redirects the \(d\) edge of \(p(x)\) to point to \(p(y)\).
\item \emph{\textbf{center} x} where \(x\) is a path, moves the center to \(p(x)\).
\item \emph{\textbf{if} x y \textbf{then} ln} where \(x,y\) are paths and \emph{ln} a line number, jumps to line \emph{ln} if \(p(x) = p(y)\) and skips to the next line if not. Line numbers can be absolute, \emph{ln}, or relative to the current line number, \emph{+ln} or \emph{-ln}.
\item \emph{\textbf{stop} message} halts the SMM, printing out \emph{message}.
\end{itemize}

The similarities highlighted above between TM and SMM are our guidelines to the specification of a generic transformation of the TM transition table to  appropriate sequences of SMM instructions.

\section{A TM-to-SMM Compiler}
\label{sec:org0da619c}
More specifically, elaborating on these similarities: each TM tape cell and each position of the head over it are represented by a node in the compiled SMM, with one dedicated direction, say \texttt{f}, pointing from head to cell and cell to head. 

The symbol on the tape cell is simply the binary representation of its index in the alphabet, using \(n\) specialized directions in the \texttt{tape} node, where \(n\) is the integer immediately superior to the base-2 log of the alphabet size. The state of the TM is encoded in a \texttt{head} node in the same way, the binary representation of its index in the state-set using \(m\) specialized directions (which may be in common with the former \(n\) ones, without confusion as a node is either a \texttt{tape} or a \texttt{head}).

A special direction in both type of nodes, say \texttt{o}, always points to a fixed initial node, the \emph{Origin}. The center of the SMM is set to the current position of the TM head, i.e. the SMM \texttt{head} node linked to the current tape cell \texttt{tape} node.

By convention, the binary representation bits in nodes are \(0\) when their direction points to self, and \(1\) when their direction points to the fixed Origin. 

Finally \texttt{tape} nodes are doubly-linked using two specialized directions, \texttt{e} and \texttt{w} (for east and west); their corresponding \texttt{head} nodes (in the \texttt{f} direction) are doubly-linked in the same way. The compiled SMM has then \(4 + max(n,m)\) directions.

The TM compiler is built as a sequence of code generation/decoration passes:

\begin{itemize}
\item \emph{State Selection}. Build a decision tree based on the state binary representation over the \(m\) bit directions, using SMM controls \texttt{if <B> o} to test each \texttt{<B>} bit direction.
\item \emph{Symbol Detection}. At each leaf of the previous tree, build a decision tree based on the symbol binary representation over the \(n\) bit directions. At each leaf of each of these new trees, add the control list implementing the \((\sigma,S)\) transition found in the TM table: binary encode the new symbol on the current \texttt{tape} node; move to the \texttt{e} or \texttt{w} \texttt{head} node as indicated in the transition (possibly creating new SMM nodes in the process, if they do not exist, hence simulating an infinite tape); finally encode the new state in this \texttt{head} cell, and make it the center. This reproduces the steps 2-4 of the TM.
\item \emph{Prologue}. Prefix the control list resulting from the previous passes with a specific control list setting up the initial tape and symbols, the initial head position and the initial state.
\end{itemize}

The compiled SMM is ran stepwise: the initial prologue to set up the TM as a first mandatory step, followed by a sequence of calls to the above transition control list. Each call to the SMM execution step implements a full transition in the TM execution.

\subsection{SMM simulation of the TM simulation of the Collatz \(3x+1\)  function}
\label{sec:orgd202022}
We exercize the TM-to-SMM compiler on the compact 3-4 TM, simulating the Collatz \(3x+1\) function, given in \cite{michel2014} and defined as:

\begin{center}
\begin{tabular}{|c|c|c|c|c|}
\hline
\((S,\sigma)\) & b(lank) & 0 & 1 & 2\\
\hline
\(A\) & bLC & 0RA & 0RB & 1RA\\
\(B\) & 2LC & 1RB & 2RA & 2RB\\
\(C\) & bRA & 0LC & 1LC & 2LC\\
\hline
\end{tabular}
\end{center}

where \(R\), \(L\) represent right and left, captured as \texttt{e}, \texttt{w} directions in the compiled SMM. This TM operates on integers represented in base 3: the symbol alphabet is \(\{b,0,1,2\}\) and the state set is \(\{A,B,C\}\). The initial tape is the base-3 representation of \(x_0\), and the initial state is set to \(A\). Note that this TM never halts as it ends up -- if the Collatz conjecture is true -- looping over the \(\{1,4,2\}\) cycle.

The following page displays twelve steps of the compiled SMM execution, starting from the initial value \(x_0=19\). On the right-hand side, top to bottom, the SMM growth is graphed with the center highlighted in gray. (The \texttt{o} direction and all binary representations of symbols and states in the SMM nodes are omitted for clarity.) On the left-hand side a more conventional TM graphical representation of the SMM computation is presented; symbols and states are shown and color-coded, with the position of the head. The initial tape cell is circled twice.

The repeated execution of the SMM control list results in values which are read (in base 3) when the TM state is \(C\) and the head is at its leftmost position, over a blank, \texttt{b}, tape cell. Note that the compact Collatz 3-4 TM computes the only odd values in the Collatz series starting at \(x_0\), skipping over all the intermediate divisions by \(2\). Here in the trace displayed next page:

\begin{center}
\begin{tabular}{|c|c|c|}
\hline
T & Base 3 & Decimal\\
\hline
0 & 201 & 19\\
7 & 1002 & 29\\
\ldots{} & \ldots{} & \ldots{}\\
\hline
\end{tabular}
\end{center}

Of course the delay between successive printings of the iterated values on the tape lengthens as the tape is progressively expanded right, forcing a longer trip of the head back to its leftmost position in state \(C\).

This compiled SMM has 6 directions, 2 of which are used for the binary representation of states and symbols. At any given time it counts \(2n+1\) nodes, where \(n\) is the length of the tape. The prologue control list is 43 lines long; the transition control list is 398 lines long (including compiler-generated comments).

(Listings for the compiled SMM program, and other Turing Machines simulations is on-line at \url{https://github.com/CRTandKDU/SMM}.)

\begin{figure}[htbp]
\centering
\includegraphics[height=18cm]{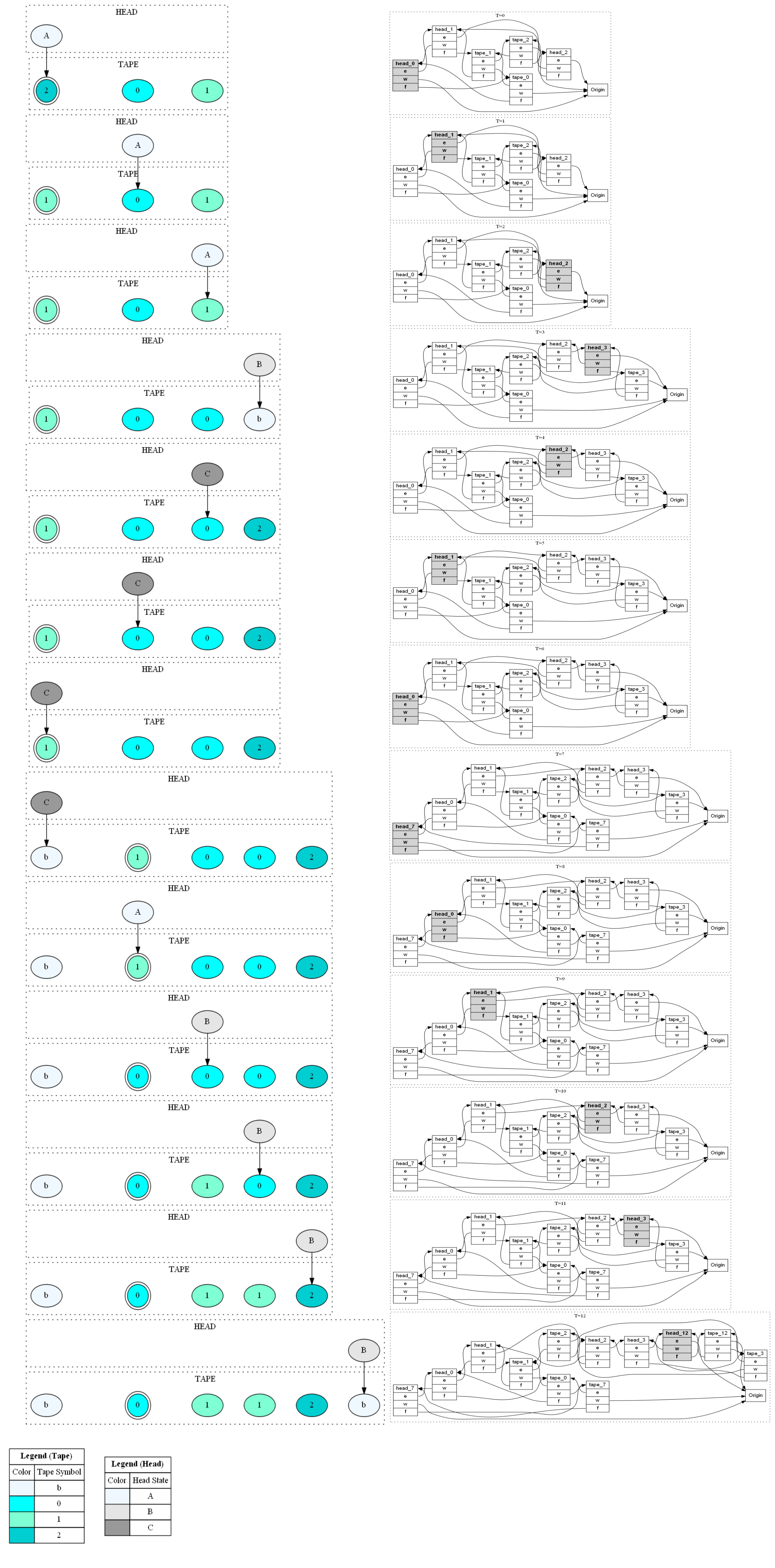}
\caption{Twelve steps of the SMM compiled from the compact Collatz 3-4 TM.}
\end{figure}
\newpage

\section{Conclusion}
\label{sec:org960c889}
Schönhage's Storage Modification Machines can generally simulate Turing Machines. This paper provides an alternate construction of such SMM simulations which finds its use in a Turing Machine to SMM compiler . Related questions on the minimal (space) complexity SMM required for simulation of complex Turing Machines may then be addressed by looking into optimizing this simple TM-to-SMM compiler.

\bibliographystyle{plain}
\bibliography{smm}
\end{document}